\documentclass[a4paper]{article}
\usepackage{indentfirst}
\usepackage{amsmath,amssymb}
\usepackage{bm}
\title{Energetics of a fluid under the Boussinesq approximation}
\author{Kiyoshi Maruyama\\
                \footnotesize Department of Earth and Ocean Sciences,
                \footnotesize National Defense Academy,\\
                \footnotesize Yokosuka, Kanagawa 239-8686, Japan}
\date{\footnotesize \today}
\begin{document}
\maketitle
\begin{abstract}
 This paper presents a theory 
 describing the energy budget of a fluid
 under the Boussinesq approximation: 
  the theory is developed in a manner 
 consistent with the conservation law of mass. 
 It shows that no potential energy is available 
 under the Boussinesq approximation, 
 and also reveals that the work done by the buoyancy force 
 due to changes in temperature corresponds to 
 the conversion between kinetic and internal energy. 
 This energy conversion, however, 
 makes only an ignorable contribution 
 to the distribution of temperature 
 under the approximation. 
 
 The Boussinesq approximation is, 
 in physical oceanography,  extended so that 
 the motion of seawater can be studied. 
 This paper considers this extended approximation as well. 
 Under the extended approximation, 
 the work done by the buoyancy force due to changes in salinity 
 corresponds to the conversion between 
 kinetic and potential energy.  
 It also turns out 
 that  the conservation law of mass 
 does not allow the condition 
 $\nabla\cdot\bm{u}=0$ on the fluid velocity $\bm{u}$ 
 to be imposed under the extended approximation; 
 the condition to be imposed instead is presented. 
\end{abstract}
\section{Introduction}
The Boussinesq approximation is frequently used 
to study the motion of a fluid 
with a non-uniform temperature distribution. 
Under this approximation, Winters et al.\ (1995) 
presented a theoretical framework for analyzing 
the energy budget of a density-stratified flow. 
Since then, in physical oceanography, 
the energetics of the overturning circulation 
of the oceans has been studied within this theoretical 
framework (Huang 1998; Hughes, Hogg \& Griffiths 2009). 
This framework, however, 
is physically inadmissible as explained below. 

Winters et al.\ assumed in their paper that the density $\rho$
appearing in their definition of the potential energy of a fluid 
fulfilled the following equation: 
\begin{equation}
 \label{1.10}
 \partial\rho/\partial t
 +\bm{u}\cdot\nabla\rho
 =-\nabla\cdot\bm{j}_{\text{d}},
 \end{equation}
where $\bm{u}$ is the fluid velocity 
satisfying the condition $\nabla\cdot\bm{u}=0$, 
and $\bm{j}_{\text{d}}$ denotes 
the diffusion flux density of mass 
allowed in their paper to exist. 
This equation can be rewritten, using $\nabla\cdot\bm{u}=0$, 
in the form 
\begin{equation}
 \label{1.20}
 \partial\rho/\partial t
 +\nabla\cdot(\rho\bm{u}+\bm{j}_{\text{d}})=0.
 \end{equation}
It follows from this expression 
that $\rho\bm{u}+\bm{j}_{\text{d}}$ is 
the momentum of unit volume of fluid 
(see Landau \& Lifshitz 1987, \S\,49). 
Hence, so long as $\bm{j}_{\text{d}}\neq0$, 
we arrive at the irrational conclusion 
that $\bm{u}$ is not the momentum of unit mass of fluid. 

The origin of this irrational conclusion is as follows. 
To derive the equation of motion for a fluid 
under the Boussinesq approximation, 
the density $\rho$ of the fluid is customarily 
assumed to be given by 
(see Landau \& Lifshitz 1987, \S\,56) 
\begin{equation}
 \label{1.30}
 \rho=\rho_{0}-\rho_{0}\beta T',
\end{equation}
where $\rho_{0}$ is a constant reference density, 
$\beta$ is the thermal expansion coefficient, 
and $T'$ is the temperature of the fluid 
measured from a reference value. 
This assumption, combined with 
the condition $\nabla\cdot\bm{u}=0$, 
gives rise to the diffusion flux of mass; 
we are consequently led to the above 
irrational conclusion. 

The assumption (\ref{1.30}), however, 
should be interpreted as a mere expedient 
to derive the equation of motion 
for the following reason. 
Suppose that the fluid is 
contained in a domain with a fixed volume $V$. 
The mass $M$ of the fluid is then given by 
\begin{equation}
 \label{1.40}
 M=\rho_{0}V-\rho_{0}\beta V\overline{T'},
\end{equation}
where $\overline{T'}$ denotes 
the averaged temperature of the fluid. 
From this expression, it follows that 
the mass $M$ varies 
when the fluid is heated or cooled. 
Evidently,  this conclusion violates 
the conservation law of mass: 
the law requires, instead of (\ref{1.30}), 
that $\rho=\rho_{0}$. 

It is now evident that the definition 
by Winters et al.\ of the potential energy of a fluid 
is physically unacceptable. 
Since their theoretical framework contains no mathematical error, 
it may be used for analyzing numerical simulations 
(see e.g.\ Gayen et al.\ 2013); 
it cannot be used, nevertheless, for analyzing real flows. 
Thus the aim of this paper is 
to present a physically reasonable theory 
that can describe the energy budget of a fluid 
under the Boussinesq approximation. 
It is also extended so that the motion of seawater can be dealt with. 
\section{Energetics under the Boussinesq approximation}
Let us  consider the motion under gravity of a viscous fluid
contained in a fixed domain $\Omega$. 
We set up in the domain 
a system of rectangular coordinates $(x_{1}, x_{2}, x_{3})$ 
with the $x_{3}$-axis taken vertically upwards.
The unit vectors in the positive
$x_{1}$-, $x_{2}$-, and $x_{3}$-directions
are respectively denoted by 
$\bm{e}_{1}$, $\bm{e}_{2}$, and $\bm{e}_{3}$.
In the following, Latin indices are used to represent 
the numbers $1$, $2$, and $3$; 
the summation convention is also implied. 

As explained in \S\,1, in view of the conservation law of mass, 
the density $\rho$ of the fluid 
should be regarded under the Boussinesq approximation as constant: 
\begin{equation}
\label{2.10}
\rho=\rho_{0}.
\end{equation}
This is consistent with the usual assumption under the approximation 
that the velocity $\bm{u}$ of the fluid is solenoidal: 
\begin{equation}
\label{2.20}
 \nabla\cdot\bm{u}=0.
\end{equation}
Nevertheless, 
the thermal expansion coefficient $\beta$ 
of the fluid is assumed not to vanish: 
\begin{equation}
\label{2.30}
 \beta=v^{-1}(\partial v/\partial T)_{p}\neq0.
\end{equation}
Here $v=1/\rho$ denotes the specific volume of the fluid;
$T$ and $p$ are respectively the temperature and 
the pressure of the fluid. 

We first wish to find an equation for 
the rate of change of the internal energy of the fluid. 
This can be derived from the general equation of heat transfer
(see Landau \& Lifshitz 1987, \S\,49): 
\begin{equation}
\label{2.40}
 \rho T\frac{Ds}{Dt}
 =\tau_{ij}\frac{\partial u_{i}}{\partial x_{j}}
 -\nabla\cdot\bm{q},
\end{equation}
in which $D/Dt$ denotes the material derivative, 
$s$ is the specific entropy of the fluid, 
$\tau_{ij}$ are the components of the viscous stress tensor, 
$u_{i}$ are those of $\bm{u}$, 
and $\bm{q}$ is the heat flux density; 
the first term on the right represents 
the heating due to viscous dissipation. 

Under the Boussinesq approximation, 
the temperature and the pressure are assumed 
to vary only slightly in the fluid (see Landau \& Lifshitz 1987, \S\,56). 
Accordingly, $T$ can be written in the form 
\begin{equation}
\label{2.50}
T=T_{0}+T',
\end{equation}
where $T_{0}$ is a constant reference temperature, 
and $T'$ is the small variation from $T_{0}$. 
We also write, denoting by $p'$ the small perturbation pressure, 
\begin{equation}
\label{2.60}
p=p_{0}+p'.
\end{equation}
Here $p_{0}$, which satisfies the hydrostatic equation, 
is defined by 
\begin{equation}
\label{2.70}
 p_{0}=-\rho_{0}gx_{3}+\mbox{constant},
\end{equation}
where $g$ is the acceleration due to gravity. 

Now let us express $Ds/Dt$ in terms of  $T$ and $p$: 
\begin{equation}
\label{2.80}
 \frac{Ds}{Dt}=\left(\frac{\partial s}{\partial T}\right)_{p}\frac{DT}{Dt}
 +\left(\frac{\partial s}{\partial p}\right)_{T}\frac{Dp}{Dt}.
\end{equation}
Here $(\partial s/\partial T)_{p}$ and $(\partial s/\partial p)_{T}$ 
are given by (see e.g.\ Batchelor 1967, \S\,1.5) 
\begin{equation}
\label{2.90}
\left(\frac{\partial s}{\partial T}\right)_{p}
 =\frac{C_{p}}{T},
 \quad
 \left(\frac{\partial s}{\partial p}\right)_{T}
 =-\left(\frac{\partial v}{\partial T}\right)_{p}
 =-v\beta,
\end{equation}
in which $C_{p}$ is the specific heat at constant pressure. 
Since $T$ and $p$ vary only slightly, 
$C_{p}$ and $\beta$ may be taken to be constant; 
we may also put $T=T_{0}$ in (\ref{2.90}). 
Thus we have, since $v=v_{0}=1/\rho_{0}$, 
\begin{equation}
 \label{2.100}
 \frac{Ds}{Dt}=\frac{C_{p}}{T_{0}}\frac{DT'}{Dt}
 -v_{0}\beta\left(\frac{Dp_{0}}{Dt}+\frac{Dp'}{Dt}\right),
\end{equation}
where (\ref{2.50}) and (\ref{2.60}) have been used. 

It is also possible to express $Ds/Dt$ in terms of 
the specific energy $e$ and the specific volume $v$: 
approximating $T$ and $p$ by $T_{0}$ and $p_{0}$, 
we get 
\begin{equation}
 \label{2.110}
 \frac{Ds}{Dt}=\frac{1}{T_{0}}\frac{De}{Dt}
 +\frac{p_{0}}{T_{0}}\frac{Dv}{Dt}.
\end{equation}
We note, however, that $Dv/Dt=v\nabla\cdot\bm{u}=0$. 
Hence it follows that 
\begin{equation}
 \label{2.120}
 \frac{Ds}{Dt}=\frac{1}{T_{0}}\frac{De}{Dt}.
\end{equation}

Using (\ref{2.10}), (\ref{2.50}), (\ref{2.100}), and (\ref{2.120}), 
we find 
\begin{align}
 \label{2.130}
 \rho T\frac{Ds}{Dt}
 &=
 \rho_{0}T_{0}\frac{Ds}{Dt}+\rho_{0}T'\frac{Ds}{Dt}\notag\\
 &=
 \rho_{0}\frac{De}{Dt}
 +\rho_{0}T'\left\{\frac{C_{p}}{T_{0}}\frac{DT'}{Dt}
 -v_{0}\beta\left(\frac{Dp_{0}}{Dt}+\frac{Dp'}{Dt}\right)\right\}.
\end{align}
Since $Dp_{0}/Dt=-\rho_{0}gu_{3}$, 
 we obtain, to the first order of primed variables, 
 \begin{equation}
 \label{2.140}
  \rho T\frac{Ds}{Dt}
  =\rho_{0}\frac{De}{Dt}
  +\rho_{0}\beta T'gu_{3}.
 \end{equation}
The substitution of this result into (\ref{2.40}) yields 
\begin{equation}
\label{2.150}
 \rho_{0}\frac{De}{Dt}
 = \tau_{ij}\frac{\partial u_{i}}{\partial x_{j}}
 -\nabla\cdot\bm{q}
 -\rho_{0}\beta T'gu_{3}.
\end{equation}
Integrating (\ref{2.150}) over $\Omega$, 
whose boundary is denoted by $\Sigma$, we have 
\begin{equation}
 \label{2.160}
 \frac{d}{dt}\int_{\Omega}\rho_{0}edV
 = \int_{\Omega}\tau_{ij}\frac{\partial u_{i}}{\partial x_{j}}dV
 -\int_{\Sigma}\bm{q}\cdot\bm{n}dS
 -\int_{\Omega}\rho_{0}\beta T'gu_{3}dV.
\end{equation}
Here $\bm{n}$ is the unit outward normal on $\Sigma$, 
and we have assumed that $\bm{u}\cdot\bm{n}=0$ on $\Sigma$. 
This is the desired equation for the rate of change 
of the internal energy. 

Let us next consider the kinetic energy of the fluid. 
Under the Boussinesq approximation, 
the equation of motion is given 
by (see Landau \& Lifshitz 1987, \S\,56)
\begin{equation}
 \label{2.170}
 \rho_{0}\frac{D\bm{u}}{Dt}
 =-\nabla p'
 +\frac{\partial\tau_{ij}}{\partial x_{j}}\bm{e}_{i}
 +\rho_{0}\beta T'g\bm{e}_{3}.
\end{equation}
From this equation, we can obtain the following equation 
for the rate of change of the kinetic energy: 
\begin{equation}
 \label{2.180}
 \frac{d}{dt}\int_{\Omega}\tfrac{1}{2}\rho_{0}|\bm{u}|^{2}dV
 =\int_{\Sigma}u_{i}\tau_{ij}n_{j}dS
 -\int_{\Omega}\tau_{ij}\frac{\partial u_{i}}{\partial x_{j}}dV
 +\int_{\Omega}\rho_{0}\beta T'gu_{3}dV,
\end{equation}
where $n_{j}$ are the components of $\bm{n}$. 

On the other hand, the potential energy of the fluid is invariable: 
\begin{equation}
 \label{2.190}
 \frac{d}{dt}\int_{\Omega}\rho_{0}gx_{3}dV=0.
\end{equation}
Hence, under the Boussinesq approximation, 
no potential energy is available. 

Adding (\ref{2.160}), (\ref{2.180}), and (\ref{2.190}), 
we get 
\begin{equation}
 \label{2.200}
 \frac{d}{dt}\int_{\Omega}\rho_{0}
 \left(\tfrac{1}{2}|\bm{u}|^{2}+gx_{3}+e\right)dV
 =\int_{\Sigma}u_{i}\tau_{ij}n_{j}dS
 -\int_{\Sigma}\bm{q}\cdot\bm{n}dS.
\end{equation}
This expression states that the total energy of the fluid 
changes owing to 
the work done by the viscous force acting on $\Sigma$ 
and owing also to the heat transfer across $\Sigma$. 
We have thus obtained a physically reasonable result. 

We now focus our attention on the equation (\ref{2.180}). 
Of the three terms on its right-hand side, 
the last term represents the work 
done by the buoyancy force: 
this force is represented by the last term of (\ref{2.170}). 
The last term of (\ref{2.180}) appears, 
with an opposite sign, 
in (\ref{2.160}) too. 
Hence the work done by the buoyancy force 
is seen to correspond to the conversion 
between kinetic and internal energy. 

Finally, let us return to the equation (\ref{2.150}). 
Rewriting this equation 
with the help of (\ref{2.100}) and (\ref{2.120}), we obtain 
\begin{equation}
\label{2.210}
 \rho_{0}C_{p}\frac{DT'}{Dt}
 = \tau_{ij}\frac{\partial u_{i}}{\partial x_{j}}
 -\nabla\cdot\bm{q}
 +\left\{\beta T_{0}\left(\frac{Dp_{0}}{Dt}+\frac{Dp'}{Dt}\right)
 -\rho_{0}\beta T'gu_{3}\right\}.
\end{equation}
The sum of the terms in the braces, 
however, is neglected 
under the Boussinesq approximation together with 
the term representing the heating due to 
viscous dissipation (see Landau \& Lifshitz 1987, \S\,56). 
Thus we have 
\begin{equation}
\label{2.220}
 \frac{DT'}{Dt}
 =-\frac{1}{\rho_{0}C_{p}}\nabla\cdot\bm{q}.
\end{equation}
This equation, combined with e.g.\ Fourier's law, 
determines the distribution of temperature 
under the Boussinesq approximation. 
However, as is apparent from the derivation, 
this temperature equation takes no account of 
the contribution of the conversion between 
kinetic and internal energy discussed above. 
\section{Energetics under the oceanographic Boussinesq approximation}
In physical oceanography, the Boussinesq approximation is 
extended to study the motion of 
seawater (see e.g.\ McWilliams 2006, \S\,2.2). 
Our aim in this section is to investigate 
the energy budget of a fluid 
under this extended approximation 
which is called in this paper 
the oceanographic Boussinesq approximation. 

We consider the same situation as that in \S\,2, 
but the fluid to be studied is now regarded as seawater. 
The salinity $c$, the mass of dissolved salt 
per unit mass of fluid, 
is assumed to change only slightly in the fluid. 
Thus we write 
\begin{equation}
 \label{3.230}
 c=c_{0}+c',
\end{equation}
where $c_{0}$ is some constant reference salinity, 
and $c'$ is the small deviation from $c_{0}$. 
The equation for $c$ is 
as follows (see Landau \& Lifshitz 1987, \S\,58): 
\begin{equation}
 \label{3.240}
 \rho\frac{Dc}{Dt}=-\nabla\cdot\bm{i},
\end{equation}
where $\bm{i}$ denotes the diffusion flux density of salt; 
it should be noted that the flux density of salt and that of water 
are given by $\rho c\bm{u}+\bm{i}$ and 
$\rho(1-c)\bm{u}-\bm{i}$, respectively. 
We also need to introduce the chemical potential $\mu$
of the fluid (see Landau \& Lifshitz 1987, \S\,58). 
It is regarded as a function of $T$, $p$, and $c$: 
\begin{equation}
 \label{3.250}
 \mu=\mu(T,p,c).
\end{equation}

When $c'$ is uniform throughout the fluid, 
the density $\rho$ should be equated to a constant as in \S\,2. 
This constant, however, may be a function of $c'$. 
Hence, in view of the smallness of $c'$, we put 
\begin{equation}
 \label{3.260}
 \rho=\rho_{0}+\rho_{0}\beta_{\text{s}}c',
\end{equation}
in which the haline contraction coefficient 
$\beta_{\text{s}}=\rho^{-1}(\partial \rho/\partial c)_{T,p}$ 
is regarded as a constant. 
Then, from the equation of continuity 
\begin{equation}
 \label{3.270}
 \rho^{-1}\frac{D\rho}{Dt}+\nabla\cdot\bm{u}=0,
\end{equation}
it follows that in general 
$\nabla\cdot\bm{u}\neq0$. 
Using (\ref{3.260}) and (\ref{3.270}), we can obtain 
\begin{equation}
 \label{3.280}
 \frac{d}{dt}\int_{\Omega}\rho gx_{3}dV
 =\int_{\Omega}(\rho_{0}gu_{3}+\rho_{0}\beta_{\text{s}}c'gu_{3})dV.
 \end{equation}
Here we have assumed, for simplicity, 
that $\bm{u}\cdot\bm{n}=0$ on $\Sigma$. 
This is the potential energy equation 
for the present problem. 

The internal energy equation can now be found 
from the following general equation 
of heat transfer (see Landau \& Lifshitz 1987, \S\,58): 
\begin{equation}
\label{3.290}
 \rho T\frac{Ds}{Dt}
 =\tau_{ij}\frac{\partial u_{i}}{\partial x_{j}}
 -\nabla\cdot\bm{q}
 +\mu\nabla\cdot\bm{i}.
\end{equation}
When $s$ is regarded as a function of $T$, $p$, and $c$, 
we get, in place of (\ref{2.100}), 
\begin{equation}
 \label{3.300}
 \frac{Ds}{Dt}=\frac{C_{p}}{T_{0}}\frac{DT'}{Dt}
 -v\beta\left(\frac{Dp_{0}}{Dt}+\frac{Dp'}{Dt}\right)
 -\left(\frac{\partial\mu}{\partial T}\right)_{p,c}\frac{Dc}{Dt}.
\end{equation}
Here the thermodynamic relation $(\partial s/\partial c)_{T,p}=-(\partial\mu/\partial T)_{p,c}$ 
has been used (see Landau \& Lifshitz 1987, \S\,59); 
note also that $v_{0}$ in (\ref{2.100}) has been replaced by $v$ 
in view of (\ref{3.260}). 
Since $(\partial s/\partial c)_{e,v}=-\mu/T$, 
we also have, in place of (\ref{2.120}), 
\begin{equation}
 \label{3.310}
 \frac{Ds}{Dt}=\frac{1}{T_{0}}\frac{De}{Dt}
 +\frac{p_{0}}{T_{0}}\frac{Dv}{Dt}
 -\frac{\mu_{0}}{T_{0}}\frac{Dc}{Dt},
\end{equation}
where $\mu_{0}=\mu(T_{0},p_{0},c_{0})$. 
Thus, neglecting terms above the first order 
in primed variables, we can obtain 
the following equation corresponding to (\ref{2.150}): 
\begin{equation}
\label{3.320}
 \rho\frac{De}{Dt}
 =-p_{0}\nabla\cdot\bm{u}
 +\tau_{ij}\frac{\partial u_{i}}{\partial x_{j}}
 -\nabla\cdot\bm{q} 
 -\rho_{0}\beta T'gu_{3},
\end{equation}
where $\rho Dv/Dt=\nabla\cdot\bm{u}$ 
and (\ref{3.240}) have been used. 
We note in addition that 
\begin{equation}
 \label{3.330}
 -p_{0}\nabla\cdot\bm{u}
 =-\nabla\cdot(p_{0}\bm{u})
 +\bm{u}\cdot\nabla p_{0}
 =-\nabla\cdot(p_{0}\bm{u})
 -\rho_{0}gu_{3}.
\end{equation}
The internal energy equation 
is therefore obtained as follows: 
\begin{equation}
 \label{3.340}
 \frac{d}{dt}\int_{\Omega}\rho edV
 =\int_{\Omega}\tau_{ij}\frac{\partial u_{i}}{\partial x_{j}}dV
 -\int_{\Sigma}\bm{q}\cdot\bm{n}dS
 -\int_{\Omega}(\rho_{0}gu_{3}+\rho_{0}\beta T'gu_{3})dV.
\end{equation}

Next, the equation of motion 
is now given by (see McWilliams 2006, \S\,2.2) 
\begin{equation}
 \label{3.350}
 \rho_{0}\frac{D\bm{u}}{Dt}
 =-\nabla p'
 +\left(\frac{\rho_{0}}{\rho}\right)\frac{\partial\tau_{ij}}{\partial x_{j}}\bm{e}_{i}
 +\rho_{0}\beta T'g\bm{e}_{3}
 -\rho_{0}\beta_{\text{s}} c'g\bm{e}_{3},
\end{equation}
in which the last term represents the buoyancy force 
due to changes in salinity, 
and the factor $\rho_{0}/\rho$ is usually identified with unity. 
Neglecting terms above the first order in primed variables, 
we can find from (\ref{3.350}) the kinetic energy equation 
\begin{equation}
 \label{3.360}
 \frac{d}{dt}\int_{\Omega}\tfrac{1}{2}\rho|\bm{u}|^{2}dV
 =\int_{\Sigma}u_{i}\tau_{ij}n_{j}dS
 -\int_{\Omega}\tau_{ij}\frac{\partial u_{i}}{\partial x_{j}}dV
 +\int_{\Omega}(\rho_{0}\beta T'gu_{3}
 -\rho_{0}\beta_{\text{s}} c'gu_{3})dV.
\end{equation}

In consequence, 
adding (\ref{3.280}), (\ref{3.340}), and (\ref{3.360}), we get 
\begin{equation}
 \label{3.370}
 \frac{d}{dt}\int_{\Omega}\rho
 \left(\tfrac{1}{2}|\bm{u}|^{2}+gx_{3}+e\right)dV
 =\int_{\Sigma}u_{i}\tau_{ij}n_{j}dS
 -\int_{\Sigma}\bm{q}\cdot\bm{n}dS.
\end{equation}
This total energy equation 
has exactly the same form as (\ref{2.200}); 
we have obtained 
a physically reasonable result again. 

The second integrand in the last term of (\ref{3.360}) 
represents the work done by the buoyancy force 
due to changes in salinity. 
Comparing (\ref{3.360}) with (\ref{3.280}), 
we observe that this work, in contrast to the work done by 
the buoyancy force due to changes in temperature, 
corresponds to the conversion 
between kinetic and potential energy. 
On the other hand, we recognize that 
the first integrand on the right-hand side of (\ref{3.280}) 
corresponds to 
the conversion between potential and internal energy; 
it appears, with an opposite sign, 
in the last term of (\ref{3.340}). 
\section{Summary and discussion}
The energy budget of a fluid has been investigated, 
in a manner consistent with the conservation law of mass, 
under the Boussinesq approximation. 
It has been shown that no potential energy 
is available under the approximation. 
It has also turned out that the work done by the buoyancy force 
due to changes in temperature corresponds to 
the conversion between kinetic and internal energy. 
This energy conversion, however, 
is altogether neglected 
in the determination of the distribution of temperature 
under the approximation. 

In contrast, under the oceanographic Boussinesq approximation, 
the work done by the buoyancy force due to 
changes in salinity represents the conversion between 
kinetic and potential energy. 
In addition, energy conversion also 
arises between potential and internal energy 
under the approximation. 

In the discussion of the oceanographic Boussinesq approximation, 
we found that in general $\nabla\cdot\bm{u}\neq0$. 
If we had assumed that $\nabla\cdot\bm{u}=0$ 
together with (\ref{3.260}), 
then the diffusion flux of mass would have arisen. 
The occurrence of this flux is physically unacceptable, as stated in \S\,1. 
Thus it follows that 
we cannot take the condition $\nabla\cdot\bm{u}=0$ 
under the oceanographic Boussinesq approximation. 

Let us examine what condition should be taken then 
in place of $\nabla\cdot\bm{u}=0$. 
The substitution of (\ref{3.230}) and (\ref{3.260}) 
into (\ref{3.240}) leads to 
\begin{equation}
 \label{4.420}
 \rho_{0}\frac{Dc'}{Dt}=-\nabla\cdot\bm{i},
\end{equation}
where terms above the first order in primed variables 
have been neglected. 
We can similarly find from (\ref{3.270}) 
the following equation: 
\begin{equation}
 \label{4.430}
 \beta_{\text{s}}\frac{Dc'}{Dt}=-\nabla\cdot\bm{u}.
\end{equation}
From (\ref{4.420}) and (\ref{4.430}), we get 
\begin{equation}
 \label{4.440}
 \nabla\cdot(\bm{u}-\beta_{\text{s}}\bm{i}/\rho_{0})=0.
\end{equation}
We have thus obtained the condition 
to be imposed in place of $\nabla\cdot\bm{u}=0$ 
under the oceanographic Boussinesq approximation; 
this condition guarantees that $\bm{u}$ is 
the momentum of unit mass of fluid.

\end{document}